\documentclass[twoside,11pt]{article}

\usepackage{obs_study_style}

\title{Enough?}

\author{\name Drew Dimmery  \email d.dimmery@hertie-school.org \\
        \addr Professor of Data Science for the Common Good\\
       Hertie School\\
       Berlin, Germany\\ 
       \AND
       \name Kevin Munger \email kevin.munger@eui.eu \\
       \addr Chair of Computational Social Science\\
       European University Institute \\
       Florence, Italy\\}

\begin{document}
\maketitle

\begin{abstract}
We respond to Aronow et al. (2025)'s paper arguing that randomized controlled trials (RCTs) are ``enough," while nonparametric identification in observational studies is not. 
We agree with their position with respect to experimental versus observational research, but question what it would mean to extend this logic to the scientific enterprise more broadly.
We first investigate what is meant by ``enough," arguing that this is a fundamentally a sociological claim about the relationship between statistical work and larger social and institutional processes, rather than something that can be decided from within the logic of statistics. For a more complete conception of ``enough," we outline all that would need to be known -- not just knowledge of propensity scores, but knowledge of many other spatial and temporal characteristics of the social world. Even granting the logic of the critique in Aronow et al. (2025), its practical importance is a question of the contexts under study. We argue that we should not be satisfied by appeals to intuition about the complexity of ``naturally occurring" propensity score functions. Instead, we call for more empirical metascience to begin to characterize this complexity. We apply this logic to the example of recommender systems developed by Aronow et al. (2025) as a demonstration of the weakness of allowing statisticians' intuitions to serve in place of metascientific data. 
Rather than implicitly deciding what is ``enough" based on statistical applications the social world has determined to be most profitable, we argue that practicing statisticians should explicitly engage with questions like ``for what?" and ``for whom?" in order to adequately answer the question of ``enough?"

\end{abstract}

\section{Introduction}

The statistical argument by \citet{aronow2025nonparametric} (ARSSS) is unimpeachable. %
The paper combines several known statistical results to show that under SUTVA, mere non-parametric identification is not sufficient to guarantee uniform consistency of an average treatment effect (additional regularity must obtain), while knowledge of the propensity score is sufficient.
It then uses simulations to show the practical value of the knowledge of propensity scores for increasing the precision of estimation; exact knowledge of propensity scores is, perhaps intuitively in hindsight, more useful when approximating propensity scores is more difficult.

This perspective cleverly ignores well-trodden ground from previous debates over whether causation requires manipulation or whether the assumption of unconfoundedness in observational causal inference can plausibly be maintained. They grant their opponents' position on these two questions. Instead, they focus on the limitations of statistical estimation arising in observational causal inference but not in RCTs.

Theirs is an immanent critique, emerging from within the logic of statistics as estimation. The hard part of statistics is not calculus, do- or otherwise. We fully agree with this point about the statistics of RCTs.

ARSSS focus on the narrower question of statistical sufficiency, but we use this as a jumping off point to consider a different type of sufficiency: the use of RCTs as one tool among others available to researchers.
But the \textit{sociological} argument implied by ARSSS is more ambitious. Their epigraph comes from the popular book by Judea Pearl, which argues that RCTs should not be ``put on a pedestal," nor should RCTs be considered ``the gold standard of causal analysis."  These concepts are not statistical but sociological, marshaled towards what can be read as an argument about the status hierarchy of research methods. ARSSS disagrees with Pearl ``insofar as a causal analysis requires drawing conclusions from actual data."   

Part of the reason for this debate over the relative esteem in which we should hold statistical methods for RCTs and observational causal inference is academic, in the pejorative sense of the word. Much more important is the pragmatic version of this question: what should social/medical scientists and statisticians \textit{do}? Should we do RCTs, or should we do observational causal inference?

To answer this question, we focus on what we think is a core contention in this debate: that ``RCTs are enough." RCTs are enough for something, surely, but enough for what? This question cannot be answered with a proposition, it must engage with a broader social context. Statistics is done in society, and so the answer to this question has to come from the sociology of science rather than from within the logic of statistics.

Our response proceeds as follows. We begin by expanding on the question of whether RCTs are ``enough": for what? This is more than a question of statistical estimation, but a broader question about the goals of science. We then argue that the standard for what is in fact ``enough" requires engaging with issues of generalizability--and that given the problem of induction, nothing can be ``enough." We then argue for the value of empirical metascience over statisticians' intuitions as a guide for issues like whether ``naturally occurring" propensity score functions tend to be simple or complex. We conclude by agreeing with ARSSS's claim that RCTs are special -- but that this specialness is not epistemological (in the form of knowledge of propensity score functions) but rather ontological (they are the only research method that creates a counterfactual world).

\section{``Enough" is a Vibe}

\cite{munger2023temporal} describes the ``agnostic impulse" currently operating in applied statistics of causal inference, which ``aims to limit the role of human subjectivity in social science." %
The agnostic impulse is the idea that one should rest the validity of scientific conclusions as little as possible on assumptions about how the world behaves, and as much as possible on simple statements about what the researcher did. The success of the agnostic impulse has created a status hierarchy in which research is preferred the fewer (non-SUTVA) assumptions it makes about the world.

But we argue that RCTs and all causal inference methods already make strong ontological assumptions when they parameterize the world into two distinct states: treatment and control\footnote{This fundamental concern is not ameliorated by the usage of multi-valued or continuous treatments, which merely allow slightly more complex operationalizations.}. Why should any of these methods be considered non-parametric?

Along the same lines, \citet{wasserman2006all} proposes that a better term for ``non-parametric statistics'' might be ``infinite dimensional statistics''.
As ARSSS demonstrate, there are differences in assumptions inherent in ``non-parametric'' methods: a method which works for all Lipschitz continuous functions is non-parametric, but the Lipshitz assumption is, itself, a serious scope condition on applicability.
In other words, it entails some form of a parametric assumption on the world.

Parametric statistics requires imposing some structure on the data. We don't like this; our structure could be wrong, we prefer to let the data speak. 

RCTs, in contrast, require imposing some structure on \textit{the world}. We cannot just let the world speak, because the language in which it speaks is too complex for us to understand.\footnote{This framing is inspired by the historian and philosopher of science Alexandre Koyr\'e in \textit{Metaphysics and Measurement}~\citep{koyre1968metaphysics}: ``Experimentation is the methodological interrogation of nature, an interrogation which presupposes and implies a \textit{language} in which to formulate the questions, and a dictionary which enables us to read the answers." The language to which he is referring is Galileo's classical mechanics; unfortunately for statisticians but fortunately for advocates of free will and self-determination, the language of the social world is far more complex than that of the flight of cannonballs.} The first step in conducting an RCT is to simplify the world to the point where treatments can be formally randomized and outcomes quantitatively measured. At this point, we no longer need to make any assumptions -- conducting the RCT performs our model of the world.\footnote{In the sense of ``performativity" developed by \cite{austin1975things}. }

This, in addition to the advantages in estimation noted by ARSSS, is hidden by the term "non-parametric". In observational work, an ontology is assumed of the data, while in an RCT the ontology is performed in/on the world.

Framing the epistemic value of RCTs this way provides a useful point of view on the problem of generalizability. RCTs can only provide their special knowledge of propensity scores if the researcher or some allied actor has \textit{control} of a given context. From ARSSS: ``In many cases, the only way to ensure that the propensity score function is sufficiently simple is to control the assignment of treatment oneself, effectively running an RCT."

By this logic, the only way to retain the agnostic status of RCTs while generalizing their results is to \textit{control the world} -- or at least, to control the parts of the world to which the results of the RCT are to be generalized.
The strength of RCTs for science is therefore also a reason for caution, politically and socially.
Making ontological choices about what ``treatment'' \textit{is} provides a clarity that is often difficult in observational settings where no such ontological control is present.
Nevertheless, an RCT is an exercise of power and should therefore be exercised responsibly.

We now discuss the implications of this observation for the standard scientific goal of theory-testing.

\section{Total Error; or, Does The World Have Enough Shoe Leather?}

ARSSS discusses conditions necessary for estimating causal effects from data.
While generic tools based around non-parametric identification suggest that there is nothing ``special'' about RCTs, the knowledge of the propensity score allows the researcher to remove regularity conditions that are required.
In short, regularity conditions place scope conditions on the world.
They say ``if the world doesn't follow our assumptions, we cannot guarantee that we will learn a causal effect.''
Of course, if these conditions hold (along with non-parametric identification), then estimated propensity scores converge to the true propensity scores, and causal effects may be learned.
RCTs, because of associated ``shoe-leather,'' bypass this necessity by knowing the propensity score by design~\citep{freedman1991statistical}.
ARSSS demonstrate this with an adversarial example where function complexity increases with sample size.%

RCTs are enough to eliminate error in estimating the SATE given a propensity score.
This, however, is not sufficient for the \emph{scientific} goal of creating generalizable theoretical knowledge.
Supposing a standard RCT, it would provide causal evidence on one sample ($S$) from a population at one site ($P$) for one realization of a theory ($R$) at one time ($T$).
Oversimplifying, then, the estimand of an RCT can be said to be $E[\tau(S, P, R, T) \mid S = s, P = p, R = r, T = t]$, where $(s, p, r, t)$ represent parts of the study design which were chosen according to some sampling process (and thus each component has some analog to a propensity score), but from which we only typically observe a single realization.
The expectation is over the probability distribution of $(S,P,R,T)$.
The RCT provides agnostic inference on this particular estimand, but it has, unfortunately, little bearing on the larger scientific endeavor, which requires control on (at least) these other source of errors.
With the notable exception of $S$, these properties of a study are not generally thought of as random, but to construct design-based solutions to generalizing over these properties, we tend to behave as if they are.
This is most common in the way standard statistical approaches to generalization construct propensity score-like quantities~\citep[e.g.][]{stuart2011propensity,hartman2015sate}.

We can imagine study designs which provide agnostic inference on other dimensions.
For instance, \citet{dahabreh2021study} proposes random sampling of sites.
This would then eliminate the site selection bias by using the site selection probabilities~\citep{allcott2015site,vivalt2020much} if we knew the probability of selection of a given site over all possible sites for a study ($P$): a shoe leather solution.
We also have a long history in survey sampling of random sampling of units from a population, which, assuming perfect rates of response from a random sample of the population (a sadly vanishing reality), would provide selection probabilities and, thus, allow us to eliminate another source of error ($S$) through shoe leather.
These are still insufficient for the scientific task, however, unless we can understand the probability that a particular instantiation of treatment (and control) was chosen over all possibilities conditional on the given theory we seek to falsify ($R$). In what sense, then, is the RCT unique?
Random sampling over sites followed by a non-parametrically identified study on the chosen sites would likewise allow agnostic inference on one of these sources of error (while leaving the other errors subject to the uncomfortable necessity of some kind of assumptions or parametric modelling).

Here we have proposed a few different dimensions necessary for some form of scientific generalization, but they are hardly exclusive.
In any data analysis there are innumerable decisions about design, operationalization, and specific analytical approaches~\citep[c.f.][]{gelman2013garden}.
If these choices are not specified clearly and if a probability distribution over the possible choices is not assigned, we cannot use shoe leather to generalize over them without constraining assumptions of their ignorability.
Of particular concern is the problem of temporal validity (generalization over $T$): the choice over when to run a given study~\citep{munger2023temporal}.
In contrast to the other sources of error we discuss, this one is distinguished by the fact that shoe-leather \emph{cannot be a solution}.
We cannot decide to run an experiment before that experiment is conceived (so the probability of running that experiment is 0\% until it is conceived), and it is implausible to imagine a mechanism which ensures that a particular experimental design will be run between the years 2200 and 2210 with a (say) 5\% probability.
How could we possibly guarantee such an eventuality?
We will almost certainly be dead, and we cannot guarantee that any institution will retain a desire to follow through on our intentions.

The thrust of our point is analogous to that in \cite{yarkoni2022generalizability}'s influential article on ``The Generalizability Crisis" in psychology experiments. In that case, standard statistical practice is to model research participants with random effects. Yarkoni argues that the same logic should apply to every element of the experiment: that we should treat every aspect of the implementation as simply one draw of a population of possible implementations. We read Yarkoni as arguing that random sampling of units is ``not enough." We would also need random sampling of every aspect of the experimental design. For information-theoretic reasons and the curse of dimensionality, the practice of experimental, quantitative social psychology can \textit{never} be enough.

Yarkoni's critique was aimed at social psychology experiments conducted in the lab. This is already an easy case: many mechanical causes have already been controlled (by the walls, lighting and temperature controls) and where the social context is restricted. For RCTs taking place in the less-controlled ``real world," the dimensionality problem is dramatically more severe.

\section{The Epistemology of Propensity Scores}

The following passage, from the second paragraph of the Discussion, is the crux of the metatheoretical argument in ARSSS: ``The question ultimately boils down to how well-behaved we believe the propensity score function to be." 

For clarity, we highlight the  sentences which lay out the argument that observational causal inference is not enough: 

``\textbf{(1) Our experience is that naturally occurring propensity score functions may not be easily characterized}, often involving many covariates interacting in complex ways.  Knowledge of the propensity score can enormously beneficial in such settings. Researchers using observational data might for this reason want to restrict their focus to settings \textbf{(2) in which the propensity score function is known to be simple. It is often difficult to discern whether this is the case.}  In many cases, \textbf{(3) the only way to ensure that the propensity score function is sufficiently simple is to control the assignment of treatment oneself}, effectively running an RCT."

(1), taken literally, is a modest and unfalsifiable claim about the authors' experience. But in the context of the rest of the section, it is clear that the authors implicitly generalize from their experience; their experience is meant to be representative of some set of relevant experiences that statisticians might have. This implicit generalization does not have any explicit empirical support. As proponents of empirical metascience, we think it would be valuable to provide some data about the distribution of ``naturally occurring propensity score functions," assuming by this that the authors mean something like ``propensity score functions in datasets used in social science not associated with an RCT." %

Already, of course, the set of actually-existing datasets is highly selected from the (much) larger set of ``naturally occurring propensity score functions." There are deep and unresolved questions in the philosophy of statistics which involved in conceptualizing what this larger set \textbf{is}.

More concretely for this paper, (1) implies that it is relatively straightforward to know (through experience) the complexity of naturally occurring propensity score functions. This contradicts the premise of (2).

This is why we need empirical metascience. The contradictory intuitions advanced in (1) and (2) about the ease with which the analyst can know whether the propensity score is simple or complex are both plausible. We believe that the statistical argument of ARSSS deserves to be taken seriously. 
To do so requires studying its applicability rigorously. We must do empirical metascience to begin to advance beyond intuition and plausibility. %
Although we lack both philosophical grounding for conceptualizing the set of naturally-occurring propensity score functions and sufficient empirical data to test our intuitions against ground-truth examples of these propensity score functions, we can at least generate data about the distribution of beliefs among statisticians about which actually-studied propensity score functions tend to be simple or complex. This data could help move us, collectively, towards either (1) or (2).

(3) returns to our discussion in Section 2. Given that we cannot speak the language of the world, the only way to converse with the world is to \textit{control} it, to force it to speak our language.

We don't want to take a side in longstanding debates about whether causation is possible absent manipulation -- but we do want to clarify how exactly the current claim about the knowledge of propensity scores maps onto that debate. The maxim of one camp is ``no causation without manipulation." One read of ARSSS that we endorse is ``no causation without \textit{control}," where \textit{control} is used (as above) in the sense of ``control theory" rather than ``control group."

To continue to think about what it means for a propensity score to be ``known," we explore the example of recommender systems discussed in ARSSS.

\section{The Case of Recommender Systems}

The primary example developed by ARSSS is that of the now-ubiquitous recommender systems deployed by tech companies to study user experiences. 

We agree with the precise point they raise about the value of recording and sharing the propensity scores used in recommender systems in order to take advantage of the statistical guarantees developed in the paper. Indeed, we endorse the broader principle that tech companies should be sharing data with the public. But from this perspective, we're not sure that we would start with propensity scores as the most valuable data to be shared. 

The propensity scores are only useful for actors which can ``speak the language" internal to the recommender systems. What would it mean to receive \texttt{XUserThreshold03: 20\%, XUserThreshold10: 80\%} for a list of 1,000,000 userids which do not map to any observable characteristic of users that we can externally observe? These would be complete propensity scores, but they would certainly not be ``enough." Other precise information must be known as well to make these scores useful as well as ``known'': treatment definitions, mappings of users to some known reality. The specific definition of treatment, for example, may be much more useful to researchers than the propensity scores themselves!

On the other hand, it is common in technical systems for a complex model to produce a ``propensity score" to propose an action for the system to take, followed by ``business logic'' which determines whether that action is safe according to ad hoc rules developed by engineers. Such a setup would provide propensity scores, but they alone would not determine treatment; without appealing to the logic of instrumental variables (and the associated strict assumptions), this could not on its own identify the causal effect of the action. Even \emph{with} the assumptions of instrumental variables, \emph{some} causal effect might be identified, but perhaps not the one of interest due to its locality. Once again, there exists a natural experiment in the recommender system, but we find ourselves back in the morass of Section 3 where it is categorically not ``enough" to answer a scientific question.

While the case is relevant and important,  there's  no reason to believe that automated recommender systems are exemplary of ``natural experiments," especially if ``natural experiment" means ``the propensity score is known even if assignment mechanism is beyond the control of the researcher."

The propensity scores are only ``knowable" because the artificial space in which the recommender system is operating is \textit{not the world}. In the sense of the word developed by Herbert Simon, these ``experiments" are not natural but \textit{artificial}~\citep{simon1969sciences}. The aspect of this case which makes the propensity scores ``knowable" is that the world has already been reduced into something radically simpler: the crucial step of controlling the world was taken much earlier, by constructing the ontologically simple virtual world of digital media.  

While we agree that these ``artificially occurring" propensity scores are in-principle knowable, there is a crucial impediment to them being known: power. The statistical shorthand of referring to these propensity scores as ``known" disregards important societal questions about where the knowledge is located and who could potentially access it.

ARSSS write that ``there are few free lunches in statistics, but using knowledge of the propensity score to avoid estimation difficulties might be one of them." Yes, but like many of the free lunches eaten by practicing statisticians, this one is primarily available to those working for a handful of large tech firms.\footnote{As a conflict of interest declaration, Dimmery received several years of these lunches while employed by Meta. Munger visited the offices occasionally in this period and attempted to make up for the relative infrequency of the free lunches in volume.}

\section{Conclusion}

In the title of the paper, ARSSS asserted that RCTs should be put on a pedestal -- that RCTs are ``enough" while other studies that are nonparametrically identified by a causal diagram are ``not enough." 

We disagree -- at least insofar as the application of knowledge generated by causal analysis
requires extrapolating beyond actual data (from the past) in order to take some action (in the future). ARSSS's assertion concerns statistical estimation and inference but does not consider cross-sectional and temporal generalizability, which may involve irresolvable challenges. Even when a causal effect is measured precisely through an RCT in a single time and place, it may still not be possible to precisely generalize it to another time and place, where it would actually be applied to inform human action.

Successfully executed RCTs are special -- but to claim that what makes them special is that ``they allow for estimation and inference at parametric rates under innocuous moment conditions" is to damn them with faint praise. The true value of RCTs is ontological: they create novel states of the world.
Since that state of the world is created by the experimenter, it is known to align with reality.
Other empirical methods can only learn from extant states of the world. The extant world is only a small subset of the set of nearby possible worlds.
The power to explore such new worlds is uniquely afforded to experimentation.\footnote{Indeed, insofar as the goal is to study the world as it is, our argument implies that RCTs may be at a disadvantage compared to observational causal inference. This point is often made about, for example, international war. But all the more reason to be more precise about our goal as researchers: to study the world as it is/has been, or to study the world as it might be? Our preference is towards the latter.} 

We revisit the pragmatic question we posed in the first section: what should social scientists and statisticians \textit{do}? Here, we follow \cite{campbell1973social}'s invocation of the ``Social Scientist as Methodological Servant": our goal should be to assist our fellow citizens in achieving \textit{their} goals. 

Starting from the world as it is, which of the possible nearby worlds do we prefer? We can acquire our high-level direction through democratic oversight~\citep{beer1993designing}, but the specific details of implementation are questions that social scientists are best-suited to answer. RCTs are special because they allow us to learn about these nearby worlds; by evaluating the results of experiments, we can learn whether a given action (treatment) is in fact a step in the right direction, building on the normative legal theory of ``democratic experimentalism" to determine which direction is the ``right" one~\citep{dorf1998constitution,sabel2012experimentalist}.

RCTs are also special because the process of figuring out how to successfully implement an RCT is necessary for getting to that world from the current one. RCTs require creativity and care in research design, as they cannot be done well without detailed knowledge of the social context under study. We may never speak the language of the world, but designing and implementing an RCT requires us to pay attention to the task of translation.

Are RCTs ``enough," in a meaningful way that observational causal inference is not? Is it the case that -- unlike for observational causal inference -- we can confidently answer scientific queries relying only on the assumptions we can verify through shoe leather? No. 

Do RCTs deserve a special place in the continuous process by which modern society should scientifically learn about itself in relation to a constantly-changing world? Absolutely.

\newpage
\acks{We would like to thank Will Lowe, Tara Slough and Laura Bronner for helpful comments on this manuscript.}

\vskip 0.2in
\bibliography{references}

\end{document}